# Synergism of the dynamics of tetrahedral hydrogen bonds of liquid water

Alexander Kholmanskiy

Science Center "Bemcom", Moscow, Russia

allexhol@ya.ru, http://orcid.org/0000-0001-8738-0189

**Abstract**

We used modified Arrhenius approximations to analyze the known temperature dependences (TDs) of water microstructure parameters and its dynamic characteristics – self-diffusion (D), viscosity (η), relaxation time. The analysis of activation energies showed a significant difference in the molecular dynamics (MD) of water in the ranges 273-298 K and 300-373 K. The features of MD in the first range were associated with the metastable ice-like phase of water, in which hexagonal clusters with tetrahedral hydrogen bonds (HBs) predominate. Based on the ratios of the signs and values of the activation energies of HBs fluctuations and the parameters of the microstructure, it was assumed that fluctuations of HBs dipoles play a key role in the mechanism of resonant activation by thermal energy of consistent reactions of deformation, rupture and formation of tetrahedral HBs in water clusters. The synergism of these reactions and the interaction of the charges of the vacant acceptor and donor tetrahedral orbitals of the oxygen atom trigger at ~298 K an explosive transition of the metastable ice-like phase of water into the argon-like phase. The synergetics of water dynamics above 298 K is adequately characterized by the product Dη, from which TDs follow the activation energies of reactions that determine the form of the Stokes-Einstein relation in the temperature ranges below and above the 298 K point.

## 1. Introduction

The structure and dynamics of the network of hydrogen bonds (HBs) in liquid water determine its unique properties [1-3]. An adequate structural and energy characteristic of the HBs network is the orientational order parameter equivalent to the degree of tetrahedrality ($q$) of the supermolecular structure (SMS) of water [4-7]. The $q$ index is associated with anomalies in the properties of not only water, but also other liquids [7]. The tetrahedral geometry of the oxygen atomic orbitals allows the formation of four HBs with the closest molecules in the first coordination shell of the central molecule. Such configuration exists in hexagonal ice (Ih) and after melting features the highest possible stability and minimum energy [4]. The temperature dependences (TDs) of the physical properties of liquid water at normal pressure and at T from 273 K to ~373 K have either extrema or deviate from Arrhenius at T < 300 K [3, 8-10]. To clarify the physical nature of these anomalies, in addition to mathematical approximations of TDs of the properties of water, computer simulation of the molecular structure and intermolecular interactions in water is widely used [1, 2, 12-14]. In this case, the reliability and accuracy of the results of computer calculations significantly depends on the physical adequacy of mathematical models [3, 9, 14]. Empirical non-Arrhenius TDs at T <300 K are usually interpolated by power-law functions of temperature with arbitrary coefficients [5, 9, 12].

The Arrhenius equation has a universal character, since physicochemical processes in the general case have one or several stages and energy barriers that determine the kinetics of reactions. The total thermal effect and activation energy ($E_A$) of these reactions can be greater than zero (exo process), less than zero (endo process) and zero [3, 9, 10]. In [3, 10, 13], it was suggested that the anomalies TDs of the physical properties of liquid water are due to correlated interactions of dipoles of fluctuating hydrogen bonds (HBs) and the mechanism of resonant absorption of thermal energy quanta by individual molecules or their correlated ensembles. Comparison of $E_A$ reactions limiting TDs of the self-diffusion coefficient (D, $E_A <0$) and dynamic viscosity ($\eta$, $E_A> 0$) and the dielectric relaxation time of water ($E_A> 0$) showed that the molecular dynamics (MD) of water is based on synergism of consistent exo and endo reactions of HBs formation and destruction [8-10]. The $E_A$ values for dynamic characteristics are almost an order of magnitude higher than thermal energy (kT), and therefore its influence on the HBs dynamics begins to manifest itself mainly in the deviations of TDs from Arrhenius at T <300 K [9, 10]. For bulk density, isobaric heat capacity, compressibility, speed of sound, stationary dielectric constant ($\varepsilon_S$), and water microstructure parameters, $E_A$ values are of the same order of magnitude or significantly less than kT [3, 9, 10]. Therefore, the role of synergism in the dynamics of HBs increases significantly and on TDs of the listed characteristics, in addition to the break at point 298 K, extrema appear at points 277 K, 308 K, 315 K and 348 K [1-3, 8-10].

In [3, 8-10, 15], the anomalous behavior of TDs properties of water in the vicinity of the 298 K point was associated with a reversible dynamic phase transition of the metastable ice-like phase into argon-like water, consisting mainly of individual molecules and dimers. Anomalies in the thermodynamics of the ice-like phase and TDs in the properties of liquid water in the range of 273-298 K are mainly due to the dominance of hexagonal clusters and tetrahedral configurations in the

HBs networks. For their quantitative description, the $q$ index is used, which is calculated and estimated using various empirical parameters of water. From the Arrhenius approximations of the known TDs of the $q$ index and other parameters of the first and second coordination shells of water, one can obtain $E_A$, a comparison of which with the characteristic energies of MD will help to understand the mechanisms of SMS rearrangement of water. For this purpose, in the present work, modified Arrhenius approximations were used to approximate the known TDs properties of water.

**2. Method**

To reveal the features of water thermodynamics at the point 298 K, as well as at the extremum points of TDs, in [3, 9, 10], the Arrhenius approximation ($F_A$) was modified by separating the thermal component in TD. For this, the $F_A$-approximations were represented by the product of the power and exponential temperature functions (T):

$$F_A = \exp(\pm E_A/RT) = T^{\pm \beta} \exp(\pm E_R/RT),$$

where $\beta = 0$, 1/2 or 1 for different water characteristics, and $E_R$ is the activation energy $F_R$-approximation TD* of the kinetics of the water structure rearrangement reaction. TD* obtained by dividing TD by $T \pm \beta$, while the empirical values of $T \pm \beta$ were interpolated by the exponent $F_T = \exp(\pm E_T/RT)$. Thus:

$$F_A = F_T F_R = \exp(\pm E_A/RT) = \exp(\pm E_T/RT) \cdot \exp(\pm E_R/RT)$$

and

$$\pm E_A = \pm E_R \pm E_T.$$

This method was used to analyze the experimental and calculated TDs of water characteristics, which were somehow related to the average number of hydrogen bonds (HBs) in a water molecule or to their average tetrahedrality index $<q>$. For TDs the distances of each oxygen atom to its four nearest oxygen atoms ($r_C$, $R_{O4}$) [4, 5] and to its fifth nearest oxygen atom ($R_{O5}$) [5] we took $\beta = 1$, as well as for TD of D [10]. For mean tetrahedrality index $<q>$ $\beta=0$, as well as for TDs of $\eta$ and $\varepsilon_S$ [10]. When $\beta = 0$, $E_T = 0$ and $E_A = E_R$ are obtained.

Empirical data for TDs characteristics of water were imported from published sources. References to these materials are given in figure captions and in table. Graphs were digitalized with the use of Paint computer application, when necessary. MS Excel application was used to plot TDs and their approximations. The extent of proximity of value $R^2$ to 1 was chosen as the reliability criterion, for approximations, in various T-intervals.

## 3. Results and discussion

Graphs of TDs and their approximations are shown in Figures 1-4. Activation energies in Arrhenius exponents in kJ/mol, their values together with known data for comparison are presented in the Table.

A kink in the vicinity of 296 K of the dependences on 1/T and Arrhenius approximations for the $q$ index and the distances $R_{O4}$ and $R_{O5}$ (see Figs. 1-3) confirms the assignment of the 298 K point in the TDs of water properties to the extreme point [3, 8-10, 15]. Significant difference in activation energies before and after this point, as well as good linearity in these intervals of T of the dependences of water parameters on 1/T and their Arrhenius approximations, testifies to the reliability of the kink in the vicinity of 298 K (see Figs. 1-4 and Table). At the qualitative level, the hypothesis of the extreme nature of the break in the TDs properties of water in the vicinity of 298 K is confirmed by the $E_A$ correlations between physically related reactions (see Table). The qualitative level implies the reliability of correlations between trends in $E_A$, obtained by different methods. In this case, the spread in the measured and calculated $E_A$ values can be large [10, 11].

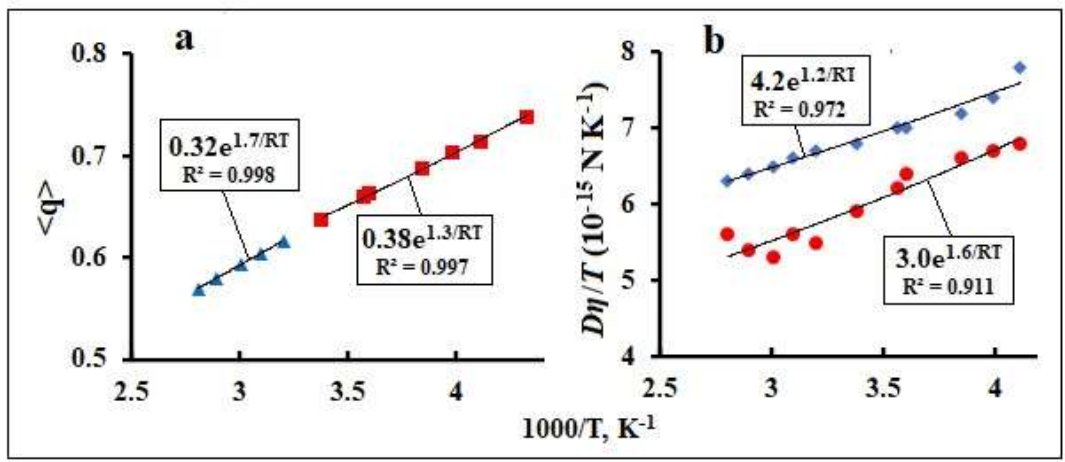

Fig. 1. (**a**) Dependence mean tetrahedrality $<q>$ on 1/T and its $F_A$-approximations. Initial data from [5]. (**b**) Dependences of composition $D\eta/T$ on 1/T and their $F_R$-approximations. Blue diamonds - the initial data were calculated in [5], red circles - the data were calculated in [5] using the formula $D = D_0[(T/T_s) − 1]^\gamma$ from [12].

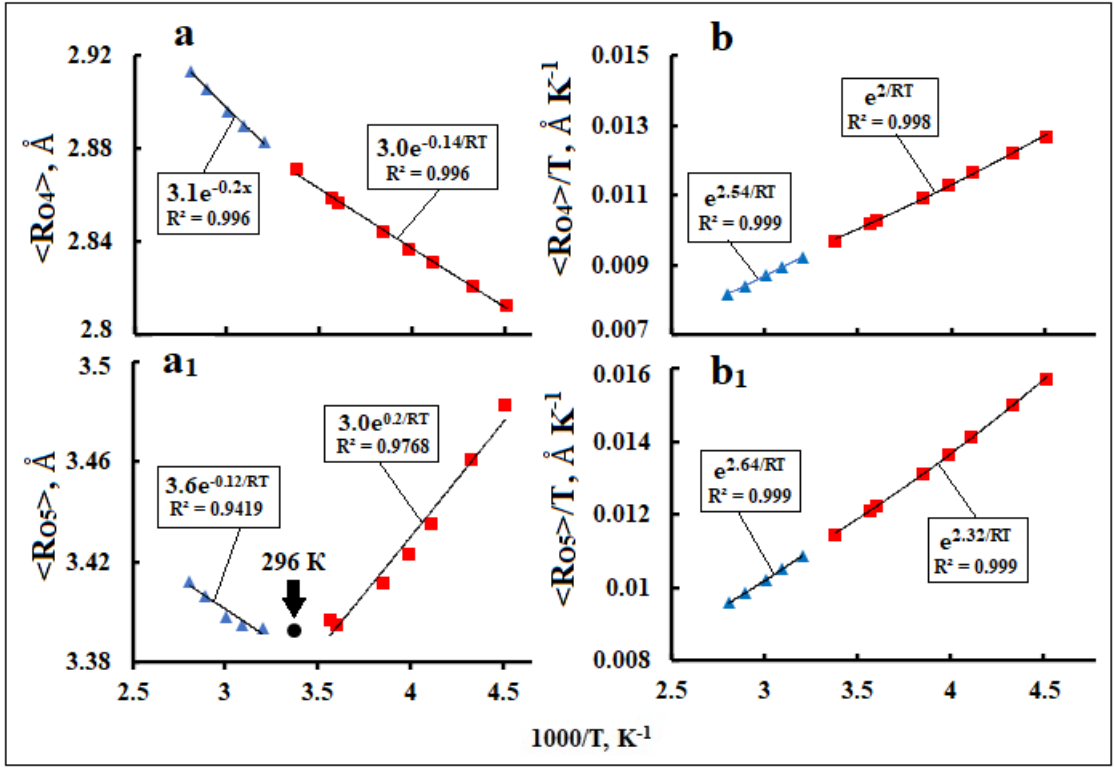

Fig. 2. Dependences of respective mean distance of each oxygen atom to its four (**a**) and fifth (**a₁**) nearest oxygen atom nearest oxygen atoms ($R_{O4}$ and $R_{O5}$) on 1/T and their $F_A$-approximations. Dependences of composition $R_{O4}/T$ (**b**) and $R_{O5}/T$ (**b₁**) on 1/T and their $F_R$-approximations. Initial data from [5].

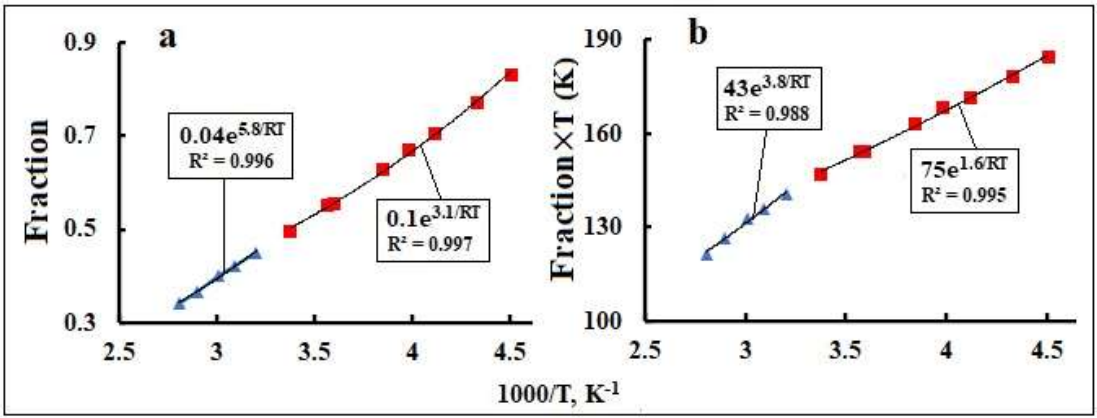

Fig. 3. (**a**) Dependence on 1/T the Fraction of OH groups that switched the acceptor through a bifurcated pathway and its $F_A$-approximation. (**b**) Dependence on 1/T the composition Fraction × T and its $F_R$-approximation. Initial data from [5].

Table 1. Temperature interval, parameter β and activation energy of Arrhenius approximations of temperature dependences of water characteristics [4, 5, 6, 10, 11, 12].

| Water characteristics | β | ΔT | $E_T$ | $E_R$ | $E_A$ ($E_R + E_T$) | Fig. [Ref] |
|---|---|---|---|---|---|---|
| | | K | | kJ/mol | | |
| <q> | 0 | 222-296 | | | 1.3 | Fig. 2; [5] |
| | | 312-356 | | | 1.7 | |
| | | 277-293 | | | 1.3 | Fig. 4; [6] |
| | | 301-363 | | | 2.7 | |
| $R_{O4}$ (<2.86> Å) | 1 | 222-296 | -2.14 | 2.0 | -0.14 | Fig. 1; [5] |
| | | 312-356 | -2.74 | 2.54 | -0.2 | |
| $R_{O5}$ (<3.44> Å) | | 222-280 | -2.12 | 2.32 | 0.2 | |
| | | 312-356 | -2.52 | 2.64 | -0.12 | |
| $r_c$ (<3.3> Å) | | 220-300 | -2.12 | 1.86 (0.999) | -0.26 (0.952) | [4] |
| $^1C_x$ (<2.9> Å) | 0 | 238-370 | | | -0.2 | [10], [11] |
| $^2C_x$ (<3.7> Å) | | 250-322 | | | 0.2 | |
| δ° | 1 | 260-360 | -2.8 | 0.4 | -2.4 | |
| $ε_s$ | -1 | 273-298 | 2.4 | 0.6 | 3.0 | [10] |
| | | 303-348 | 2.7 | 1.3 | 4.0 | |
| Fraction of bifurcation | | 222-296 | 2.2 | 1.6 | 3.8 | Fig. 3; [5] |
| | | 312-356 | 2.7 | 3.1 | 5.8 | |
| $D_n$ (N) | 1 | 243-356 | -2.4 | 1.6 | -0.8 | [5], [12] |
| | | 243-356 | -2.4 | 1.2 | -1.2 | |
| | | 273-298 | -2.4 | 1.7 | -0.7 | [10] |
| | | 300-373 | -2.7 | 0.3 | -2.4 | |

In the vicinity of 298 K, as a result of bifurcation of the structure of a tetrahedral cell with $R_{O4}$, a strained structure characterized by the parameter $R_{O5}$ can form. In addition, at 298 K, hexagonal clusters can resonantly rearrange into tetrahedral chains of six water molecules [10, 16]. With a further increase in T in the $R_{O5}$ structure and chains, the distortion and stress of tetrahedral HBs will increase, which can initiate a chain reaction of breaking one or two HBs, which will lead to the complete decomposition of the metastable ice-like phase of SMS water. The process of ice melting and crystallization of water occurs by a similar mechanism, including supercooled water [10]. The dynamic phase transition at 298 K is confirmed by the proportionality of the $E_A$ values of reactions bifurcation and generation of structures with distorted HBs at T > 298 K (see Fig. 4 and Table) to the average energy of one HB at these T (~5.4 kJ/mol) [3]. The $E_A$ values of the self-assembly of water dipoles into domains with a high $ε_s$ value have the same order (see Table), and the $E_A$ modulus of the endothermic HBs distortion reaction at T > 298 K (-11 kJ/mol, Fig. 4a) coincides with the $E_A$ TD of the rotational time relaxation of water [17].

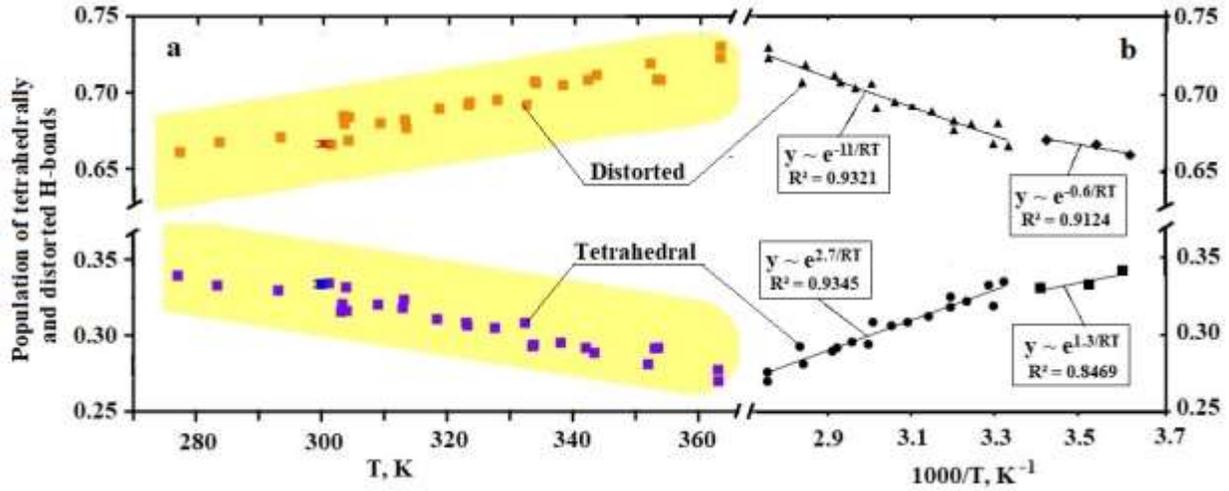

Fig. 4. Dependences of the population of tetrahedrally H-bonded species and species with distorted H-bonds on T (**a**) and on 1/T (**b**); their $F_A$-approximations (**b**). (**a**) Adapted figure from [6].

In [11], the influence of van der Waals (vdW) interactions on TDs amplitude of fluctuations in the HB angle ($\delta°$) and oxygen-oxygen radial distribution ($g_{oo}$) in the first ($^1C_X$) and second ($^2C_X$) coordinate shells of the central molecule. One variant is simulation of the tetrahedral water network corresponds to too rigid HBs, and the second variant is very weak HBs. In [10], from the data [11], the activation energies were determined, taking β for $\delta°$ and $g_{oo}$ equal to 1 and 0, respectively (see Table). In this case, the difference in $E_A$ values in the two variants was for $\delta°$ ~12%, and for $g_{oo}$ ~20%. The $F_A$-approximations for other calculated data give the same order of scatter in $E_A$ estimates. The $\delta°$ value is, in principle, related to the $q$ index, and the $g_{oo}$ values for both shells should correlate with the metric parameters $r_C$, $R_{O4}$ and $R_{O5}$. The synergism of HBs rearrangements within the first two coordination shells is evidenced by the closeness of the moduli and the opposite signs of the $E_A$ values for $\delta°$ and $q$. The magnitudes and signs of $E_A$ reactions are related in a similar way, which determine changes in the metric characteristics of the tetrahedral shells of the central oxygen atom (see Table).

From the Table it follows that the values of $E_A$ for $\delta°$ ($E_A^\delta$) and $E_R$ for $g_{oo}$ ($E_R^g$) have different signs and their moduli satisfy the relation:

$| E_A^\delta | \sim | E_R^g | \sim kT.$

On the other hand, with different signs of $E_R$ for $\delta°$ and $E_A$ for $g_{oo}$ for their modules the following relation is observed:

$$|E_R^\delta| \sim |2E_A^g| \ll kT.$$

These relations confirm the key role of fluctuations in the amplitude of the HBs in the mechanism of converting thermal energy quanta into the energy of consistent reactions of deformation and rupture of tetrahedral HBs in SMS water [10]. Thus, fluctuations of HBs dipoles and charges on vacant tetrahedral orbitals of the oxygen atom control the dynamics of water at the microlevel.

The average number of HBs in tetrahedral configurations of SMS in the range 273–298 K varies from 3 to 5 [18], and the charge density on HBs fluctuates synchronously with fluctuations $\delta°$ [19]. Thus, it can be assumed that SMS is saturated by molecules with vacancies for donor and acceptor HBs, and the medium is filled with a fluctuating electric field [16, 20]. Under these conditions, the Coulomb forces and tunneling jumps of protons [21] initiate energetically consistent rearrangements of HBs in the first two shells and rotational-translational shifts of molecules. This is confirmed by the coincidence, in the range of 273–298 K, of the activation energies TDs of the fraction of the tetrahedral structure and the average number of HBs in water [10, 18]. Both $E_A$ values are 1.3 kJ/mol (see Fig. 4b and [10]).

The product Dn has the dimension N and adequately characterizes the friction force in water. Taking this into account, it was proposed in [10] to use Dη for a quantitative assessment of the synergism of physics of self-diffusion and viscosity. From TDs of Dη, calculated in two ways in [5], one can obtain the energies of the $E_R$ reactions that are responsible for the nonlinearity of the Stokes-Einstein equation [10]:

$$D = C\frac{T}{\eta} \exp(E_R/RT).$$

In [10], from experimental TDs of D and η, the values of $E_R$ (see Table) and constants C were obtained - $3.5 \cdot 10^{-15}$ in the range of 273-298 K and $6.2 \cdot 10^{-15}$ in the range of 300-363 K. From the original calculation method [5] TDs of the composition Dη/T for T <298, the value $E_R = 1.6$ kJ/mol, which is close to $E_R$ from [10], follows. And from TDs obtained in [5] using the formula $D = D_0 [(T/T_s) - 1]^\gamma$ from [12] with arbitrary coefficients ($D_0$, $T_s$, $\gamma$), followed by a quarter lower $E_R = 1.2$ kJ/mol. In this case, both variants of calculating TDs of the composition Dη/T [5] give $E_R$ values for the interval T > 300 K 4-5 times higher than the experimental values (see Table).

## 4. Conclusions

Thus, the modified Arrhenius approximations make it possible to obtain the activation energies of the reactions responsible for the molecular physics of anomalous temperature dependences of the dynamic characteristics of water and the parameters of its microstructure. Fluctuations of the amplitudes of tetrahedral hydrogen bonds in hexagonal water clusters channel heat energy into synergistic reactions of hydrogen bond breaking and trigger a burst transition of the metastable ice-like phase of water into the argon-like phase at ~ 298 K. The synergetics of water dynamics above 298 K is adequately characterized by the product $D\eta$. From the temperature dependence of $D\eta$, the activation energies of the reactions follow, which determine the shape of the Stokes-Einstein relation in the temperature ranges below and above the 298 K point.